\documentstyle[11pt,titlepage]{article}
\begin{document}
\newcommand{\sect}[1]{\setcounter{equation}{0}\section{#1}}
\renewcommand{\theequation}{\thesection.\arabic{equation}}

\newcommand{\be}{\begin{equation}}
\newcommand{\ee}{\end{equation}}

\newcommand{\bea}{\begin{eqnarray}}
\newcommand{\ena}{\end{eqnarray}}
\newcommand{\nn}{\nonumber}

\newcommand{\ta}{{\mbox{\tiny{A}}}}
\newcommand{\tb}{{\mbox{\tiny{B}}}}
\newcommand{\tc}{{\mbox{\tiny{C}}}}
\newcommand{\td}{{\mbox{\tiny{D}}}}

\newcommand{\bz}{{\bar{z}}}
\newcommand{\bc}{{\bar{c}}}
\newcommand{\bh}{{\bar{h}}}

\newcommand{\al}{{\alpha}}
\newcommand{\bt}{{\beta}}
\newcommand{\gm}{{\gamma}}
\newcommand{\dt}{{\delta}}
\newcommand{\eps}{{\epsilon}}
\newcommand{\vep}{{\varepsilon}}
\newcommand{\vp}{\varphi}
\newcommand{\la}{{\lambda}}
\newcommand{\si}{{\sigma}}
\newcommand{\th}{\theta}

\newcommand{\thb}{\bar{\theta}}
\newcommand{\sib}{{\bar{\sigma}}}

\newcommand{\da}{{\dot{\alpha}}}
\newcommand{\db}{{\dot{\beta}}}
\newcommand{\dg}{{\dot{\gamma}}}
\newcommand{\dd}{{\dot{\delta}}}
\newcommand{\dv}{{\dot{\varphi}}}

\newcommand{\La}{\Lambda}
\newcommand{\Lab}{\bar{\Lambda}}

\newcommand{\ca}{{\cal A}}
\newcommand{\cb}{{\cal B}}
\newcommand{\cc}{{\cal C}}
\newcommand{\cd}{{\cal D}}
\newcommand{\ce}{{\cal E}}
\newcommand{\cf}{{\cal F}}
\newcommand{\ch}{{\cal H}}
\newcommand{\cl}{{\cal L}}
\newcommand{\cq}{{\cal Q}}
\newcommand{\cv}{{\cal V}}
\newcommand{\cw}{{\cal W}}

\newcommand{\bV}{\bar{V}}
\newcommand{\cwb}{{\bar{\cal{W}}}}

\newcommand{\f}[2]{{\textstyle\frac{#1}{#2}}}

\newcommand{\tr}{{\rm tr}}
\newcommand{\ie}{{\em i.e. }}

\font\fifteen=cmbx10 at 15pt
\font\twelve=cmbx10 at 12pt

\begin{titlepage}

\begin{center}

\renewcommand{\thefootnote}{\fnsymbol{footnote}}

{\twelve Centre de Physique Th\'eorique\footnote{Unit\'e Propre de
Recherche 7061 }, CNRS Luminy, Case 907}

{\twelve F-13288 Marseille -- Cedex 9}

\vspace{3 cm}

{\fifteen THE N=2 VECTOR-TENSOR MULTIPLET, \\[3mm]
          CENTRAL CHARGE SUPERSPACE, \\[3mm] 
          AND CHERN-SIMONS COUPLINGS}

\vspace{1.4 cm}

\setcounter{footnote}{0}
\renewcommand{\thefootnote}{\arabic{footnote}}

{\bf Richard GRIMM, Maximilian HASLER\footnote[1]{allocataire
M.E.S.R.} and Carl HERRMANN\footnotemark[1]}

\vspace{3 cm}

{\bf Abstract}

\end{center}

We present a new, alternative interpretation of the vector-tensor
multiplet as a 2-form in central charge superspace.
This approach provides a geometric description of the
(non-trivial) central charge transformations {\em ab initio}
and is naturally generalized to include couplings of
Chern-Simons forms to the antisymmetric tensor gauge field,
giving rise to a $N=2$ supersymmetric version
of the Green-Schwarz anomaly cancellation mechanism.                             

\vfill


\bigskip

\bigskip

\noindent June 1997

\noindent CPT-97/P. 3499

\bigskip

\noindent anonymous ftp : ftp.cpt.univ-mrs.fr

\noindent web : www.cpt.univ-mrs.fr

\renewcommand{\thefootnote}{\fnsymbol{footnote}}

\end{titlepage}
\sect{Introduction}

\indent

As its name already indicates, the vector-tensor multiplet contains among its 
bosonic components two abelian gauge potentials, a vector and an antisymmetric
tensor, subject to intriguing nontrivial central charge transformations:
the vector transforms into the dual of the fieldstrength of the antisymmetric
tensor and the antisymmetric tensor transforms into the dual fieldstrength
of the vector.

These (and other) features have been established in terms
of component fields in the original work of 
Sohnius, Stelle and West \cite{SSW80}
and rederived in the context of string compactification \cite{dWKLL95} more
recently.

As to a superspace formulation, the nontrivial structure in the 
central charge sector suggests a formulation in central charge superspace, \ie
a generalization of ordinary superspace with additional bosonic directions
corresponding to the central charges, as described by Sohnius \cite{S78,G96}.
Recall that in this language, ordinary space-time translation 
as well as supersymmetry and central charge transformations are described
as generalized translations in superspace.

The fact that the vector-tensor multiplet contains two gauge potentials
of different nature suggests that there are two possibilities to describe
it in the framework of superspace geometry.

On the one hand, as has been explained in \cite{HOW97}, one may start from 
an abelian 1-form gauge potential in central charge superspace with
judiciously chosen constraints and identify the 2-form gauge potential
in this geometric structure {\em a posteriori}.

On the other hand, one may start from a generic 2-form gauge potential in
$N=2$ central charge superspace, as will be described in the present paper,
and identify the 1-form gauge potential as a certain substructure.

As will become clear in the next chapter, 
this latter formulation provides the veritable generalization
of the well-known linear superfield 
formalism of $N=1$ supersymmetry to the case 
of $N=2$. Moreover, as will be explained in the third chapter, it allows
to incorporate central charge transformations {\em ab origine} at the
geometric level. Most importantly, this kind of superspace formulation
provides the natural setting for the coupling of Chern-Simons forms to
the antisymmetric gauge potential: in chapter {\bf 4} we recall the basic
properties of Chern-Simons forms together with their couplings to the
2-form geometry in central charge superspace. Finally, in chapter {\bf 5} we
present the complete description of the coupling of Yang-Mills Chern-Simons
forms to the vector-tensor multiplet.

\sect{2-form gauge potential in central charge superspace}

\indent

Central charge superspace \cite{G96} 
is a generalization of usual $N=2$ 
superspace\footnote{as to the conventional $N=2$ superspace without
central charge our notations follow those of ref. \cite{MM89}},
which, in addition to the vectorial and spinorial coordinates
$x^a$ and $\th {}^\al_\ta$, $\thb_\da^\ta$ contains coordinates
$z$ and $\bz$ referring to the central charge directions.
Introducing the notation 
\be Z^\ca = (x^a, \th {}^\al_\ta, \thb_\da^\ta, z, \bz), \ee
one defines the frame (vielbein) $E^\ca$ in this superspace
to have the property
\be d E^\ca \ = \ T^\ca \ = \ 
        \frac{1}{2} E^\cb E^\cc \, T_{\cc\cb} {}^\ca. \ee
As usual in superspace geometry, the constant translational
torsion is defined as
\be T {}_\gm^\tc {}^\db_\tb {\,}^a \ = \ 
    -2i \, \dt_\tb^\tc (\si^a \eps)_\gm {}^\db, \ee
whereas the non-vanishing central charge torsion components are
\be 
T {}_\gm^\tc {}_\bt^\tb {\,}^z \ = \ 
     \eps_{\gm \bt} \, \eps^{\tc \tb} c^z, \hspace{1cm}
T {}^\dg_\tc {}^\db_\tb {\,}^\bz \ = \
      \eps^{\dg \db} \, \eps_{\tc \tb} \, \bc^\bz,
\ee
with $c^z$ and $\bc^\bz$ constant and related to each other by
complex conjugation.
These constant torsions appear in the anticommutators of the covariant
spinorial derivatives, \ie
\be \{ D {}_\gm^\tc, D {}^\db_\tb \}  \ = \ 
  2i \, \dt_\tb^\tc (\si^a \eps)_\gm {}^\db \frac{\partial}{\partial x^a}, \ee

\be 
\{ D {}_\gm^\tc, D {}_\bt^\tb\} \ = \ 
- \, \eps_{\gm \bt} \, \eps^{\tc \tb} \, c^z \frac{\partial}{\partial z},
\hspace{1cm}
\{ D {}^\dg_\tc, D {}^\db_\tb \}  \ = \
- \, \eps^{\dg \db} \, \eps_{\tc \tb} \, \bc^\bz \frac{\partial}{\partial \bz}.
\ee

Given this basic superspace structure we define a 2-form gauge potential
$B$ subject to gauge transformations $B' \ = \ B + d \La$ with $\La$ a
superspace 1-form and invariant fieldstrength $H \ = \ dB$.
The geometry of the central charge superspace 3-form
\be H \ = \ \frac{1}{3!} \, E^\ca E^\cb E^\cc \, H_{\cc \cb \ca}, \ee
is constrained in such a way that all its components are expressed in terms
of one single, real, superfield $L$. This basic covariant superfield is
identified in
\be H{}_\gm^\tc {}^\db_\tb {\,}_a \ = \ 
    -2i \, \dt_\tb^\tc \ (\si_a \eps)_\gm {}^\db\  L, \ee
wheras the other nonvanishing components are given as
\be 
c^z H_z {}_\tb^\db {}_\ta^\da \ = \ 
          - 8 \, \eps^{\db \da} \eps_{\tb \ta} L, \hspace{1cm} 
\bc^\bz H_\bz {}_\bt^\tb {}_\al^\ta \ = \ 
          - 8 \, \eps_{\bt \al} \eps^{\tb \ta} L, \\[1mm]  
\ee
\be 
H {}_\gm^\tc {\,}_{ba} \ = \ 
          2 (\si_{ba})_\gm {}^\vp D {}_\vp^\tc L, \hspace{1cm}
H {}^\dg_\tc {\,}_{ba} \ = \ 
          2 (\sib_{ba})^\dg {}_\dv \, D {}^\dv_\tc L, \\[1.6mm] 
\ee
\be
c^z H_z {\,}_\tb^\db {\,}_a \ = \ 
          4i \, (\sib_a \eps)^\db {}_\bt \, D {}^\bt_\tb L, \hspace{1cm}
\bc^\bz H_\bz {\,}_\bt^\tb {\,}_a \ = \
          4i \, (\si_a \eps)_\bt {}^\db \, D {}_\db^\tb L, \\[1.8mm] 
\ee
\be
c^z \bc^\bz H_{\bz \, z} {\,}_\ta^\da \ = \ - 8 \, D {}_\ta^\da L, 
\hspace{1cm}
\bc^\bz c^z H_{z \, \bz} {\,}_\al^\ta \ = \ - 8 \, D {}_\al^\ta L, \\[2mm] 
\ee
\be
c^z H_z {\,}_{ba} \ = \ 
          (\eps \si_{ba})^{\bt \al} D_{\bt \ta} D {}^\ta_\al L, \hspace{1cm}
\bc^\bz H_\bz {\,}_{ba} \ = \ 
          (\eps \sib_{ba})_{\db \da} \, D^{\db \ta} D {}_\ta^\da L,  
\ee
As to the purely vectorial component $H_{cba}$ we introduce the dual tensor
given as $ \ 3! \, h^d \ = \ \vep^{dcba} H_{cba}$, which, in turn, 
is identified in the covariant superfield expansion of $L$ according to
\be [ D {}_\bt^\tb,  D {}_\ta^\da ] L \ = \ 
              - 2 \, \dt^\tb_\ta \, h_\bt {}^\da . \ee
More precisely, it is the trace part in the $N=2$ indices which determines
the dual of $H_{cba}$, while the traceless part in $\tb$ and $\ta$ is a
constraint equation for the superfield $L$.
In addition, the superspace Bianchi identities give rise to the constraints
\be 
\sum_{\tb \ta} D_{\da \tb} D {}_\ta^\da L \ = \ 0, \hspace{1cm}
\sum^{\tb \ta} D^{\al \tb} D {}^\ta_\al L \ = \ 0.
\ee 
The geometric structure exhibited so far clearly indicates that the 
superfield $L$ is the generalization to $N=2$ extended supersymmetry
of the $N=1$ linear superfield.

Finally, one identifies the dual of $H_{cba}$ as well in
\be c^z \bc^\bz H_{\bz \, z \ a} \ = \ 8i \, h_a. \ee
This completes our discussion of the components $H_{\cc \cb \ca}$
of the superspace fieldstrength $H=dB$, which are completely expressed
in terms of the single superfield $L$, subject to three second order
spinorial constraints.

In order to discuss the $z$ and $\bz$ dependence of this superfield we
define now
\be  h \ = \ c^z \partial_z L, \hspace{1cm} 
         \bh \ = \ \bc^\bz \partial_\bz L. \ee
As a consequence of the constraint equations on $L$ itself one finds
easily the relations
\be 
D {}^\ta_\al \bh \ = \ 2i \, \partial_\al {}^\da D {}^\ta_\da L,
\hspace{1cm}
D {}_\ta^\da h \ = \ 2i \, \partial^\da {}_\al D {}_\ta^\al L.
\ee
Moreover, applying another spinorial derivative one arrives at
\be c^z \partial_z \bh \ = \ \bc^\bz \partial_\bz h \ = \
\bc^\bz c^z \partial_z \partial_\bz L \ = \ 4 \, \Box L. \ee

Supersymmetry as well as central charge transformations will be 
obtained from covariant translations in the supercoordinates $Z^\ca$.
\be Z^\ca \mapsto Z^\ca + \zeta^\ca, \ee
with
\be \zeta^\ca \ = \ 
  \left( \zeta^a, \zeta^\al_\ta, \zeta^\ta_\da, \zeta^z, \zeta^\bz \right) \ee 
When acting on differential forms, we employ the superspace
Lie-derivative
\be L_\zeta \ = \ \imath_\zeta \, d + d \; \imath_\zeta, \ee
with the inner product acting as an antiderivative such that
$\imath_\zeta E^\ca = \zeta^\ca$.
As a consequence the 2-form gauge potential changes under the
combination of such a diffeomorphism and a 1-form gauge transformation
as
\be B \ \mapsto \ B +  L_\zeta B + d \bt \ = \ 
         B + \imath_\zeta H + d (\bt + \imath_\zeta B), \ee
and we define, as usual, covariant superspace translations such that
\be B \ \mapsto \ B + \dt B, \hspace{1cm} \mbox{with} \hspace{1cm} 
\dt B \ = \ \imath_\zeta H. \ee
This combination of a superspace diffeomorphism and a field dependent
compensating 1-form gauge transformation is customarily applied to the
description of supersymmetry transformations. When applied to central
charge transformations one simply has
\be \dt B \ = \ \imath_{\zeta^z} H + \imath_{\zeta^\bz} H. \ee
Specifying to the components $B_{ba}$ 
and $B_{za}$, $B_{\bz a}$ one simply obtains
\be \dt B_{ba} \ = \ \zeta^z H_{zba} + \zeta^\bz H_{\bz ba}, \ee
and
\be \dt B_{za} \ = \ \zeta^\bz H_{\bz z a}, \hspace{1cm}
    \dt B_{\bz a} \ = \ \zeta^z H_{z \bz a}, \ee
relations which will be useful in a short while in the context
of central charge transformations of the vector-tensor multiplet. 

\sect{The vector-tensor multiplet from 2-form geometry}

\indent

The component field content of the superspace geometry described in the
previous section is slightly more general than that of the vector-tensor
multiplet. 
Although the antisymmetric tensor gauge potential is included
manifestly, the abelian vector gauge potential remains to be identified.
For this purpose we recall that the component fields 
of the vector tensor multiplet are supposed to depend on
the central charge parameters only through the combination
$ \ z + \bz \ $.

We shall implement this particular property here in considering
the superspace 1-forms $ \ V = \imath_{c^z} B = E^\ca c^z B_{z \ca} \ $ and
$ \ \bV = \imath_{\bc^\bz} B = E^\ca \bc^\bz B_{\bz \ca} \ $ in the particular
combination $ \ A = \bV - V \ $. In view of these identifications we then
require $ \ L_{c^z} B = L_{\bc^\bz} B \ $, and obtain, as a consequence,
\be dA \ = \ d (\imath_{\bc^\bz} B - \imath_{c^z} B) 
       \ = \ \imath_{c^z} H - \imath_{\bc^\bz} H
       \ = \ F. 
      \ee
It should be clear that the reality 
condition $ \ L_{c^z} = L_{\bc^\bz} \ $ applies
to any superfield appearing in our geometry, leading, in particular
to the identification $ \ \bh = h \ $ of the superfields defined above.

At the same time this specification provides the missing link which
allows now to identify the component fields of the vector-tensor
multiplet as well as their supersymmetry and their central charge
transformations.

As to the multiplet itself, we identify, in the usual way,
component fields as lowest components of superfields, {\em viz.}
\begin{eqnarray*}
& L| \ = \ L(x), & \hspace{7mm} D {}^\ta_\al L| \ = \ \La {}^\ta_\al (x),   
\hspace{7mm} A_a| \ = \ A_a(x), \\[2mm]   
& h| \ = \ D(x), & \hspace{7mm} D {}^\da_\ta L| \ = \ \Lab {}^\da_\ta (x),
\hspace{7mm} B_{ba}| \ = \ B_{ba}(x). 
\end{eqnarray*}
Supersymmetry transformations are determined as 
\bea
\dt L &=& 
    \zeta^\al_\ta \La {}^\ta_\al + \zeta^\ta_\da \Lab {}^\da_\ta, \\[1.8mm]
\dt \La {}^\ta_\al &=& 
    \zeta {}_\da^\ta (\sib^a \eps)^\da {}_\al \, i \partial_a L
  - \f{1}{2} \, (\si^{ba})_\al {}^\bt \zeta {}_\bt^\ta \, \partial_b A_a
  - \f{1}{2} \zeta {}_\da^\ta (\sib_d \eps)^\da {}_\al
                \, \vep^{dcba} \partial_c B_{ba}
  - \f{1}{2} \,  \zeta {}_\al^\ta \, D, \\[1mm]
\dt \Lab {}^\da_\ta &=& 
    \zeta^\al_\ta (\si^a \eps)_\al {}^\da \, i \partial_a L
  + \f{1}{2} \, (\sib^{ba})^\da {}_\db \zeta {}_\ta^\db \, \partial_b A_a
  + \f{1}{2} \, \zeta^\al_\ta (\si_d \eps)_\al {}^\da 
                \, \vep^{dcba} \partial_c B_{ba} 
  - \f{1}{2} \,  \zeta {}_\ta^\da \, D, \\[1.8mm]
\dt A_a &=& - 4i \, \zeta^\al_\ta (\si_a \eps)_\al {}^\da \La {}^\ta_\da 
    + 4i \, \zeta^\ta_\da (\sib_a \eps)^\da {}_\al \Lab {}^\da_\ta, \\[1.8mm]
\dt B_{ba} &=& 2 \, \zeta^\bt_\ta (\si_{ba})_\bt {}^\al \La {}^\ta_\al 
   + 2 \, \zeta^\ta_\db (\sib_{ba})^\db {}_\da \Lab {}^\da_\ta,\\[1.5mm]
\dt D &=& 
  -2i \, \zeta^\al_\ta (\si^a \eps)_\al {}^\da \, \partial_a \Lab {}^\ta_\da
  -2i \, \zeta^\ta_\da (\sib^a \eps)^\da {}_\al \, \partial_a \La {}^\al_\ta.
\ena

Since central charge transformations are understood as translations
in $z$, $\bz$ space we shall adapt them to the 
central charge reality condition in employing the parametrization
$\zeta^z = w c^z$ and $\zeta^\bz = w \bc^\bz$.
Covariant central charge translations of the gauge potentials $A$ and $B$
are then defined as combinations of these constant diffeomorphisms
and, again, properly identified compensating field dependent
gauge transformations such that
\be \dt_{c.c.} A \ = \ \imath_{\zeta^z} F + \imath_{\zeta^\bz} F
                  \ = \ -2 w \, \imath_{c^z} \imath_{\bc^\bz} H, \ee
\be \dt_{c.c.} B \ = \ \imath_{\zeta^z} H + \imath_{\zeta^\bz} H 
                 \ = \ w \, ( \imath_{c^z} H + \imath_{\bc^\bz} H ). \ee
Projecting to the vectorial components of these superspace forms
and using previous results and identifications one finds that the
vector gauge potential transforms into the dual of the fieldstrength
of the tensor potential, whereas the tensor potential transforms into the
dual of the fieldstrength of the vector potential, in more explicit terms
\bea \dt_{c.c.} A_a &=& -8i \, w \, \eps_{abcd} \, \partial^b \! B^{cd}, \\[2mm] 
 \dt_{c.c.} B_{ab} &=& i \, w \, \eps_{abcd} \, \partial^c \! A^d. \ena
As the remaining component fields were identified as lowest components
of covariant superfields their central charge transformations are 
obtained straightforwardly to be
\bea
\dt_{c.c.} L &=& 2 \, w \, D, \\[1.5mm]
\dt_{c.c.} \La {}^\ta_\al &=& 
-4i \, w \, (\si^a \eps)_\al {}^\da \, \partial_a \Lab {}^\ta_\da, \\[1.5mm]
\dt_{c.c.} \Lab {}^\da_\ta &=& 
-4i \, w \, (\sib^a \eps)^\da {}_\al \, \partial_a \La {}^\al_\ta, \\[1.5mm]
\dt_{c.c.} D &=& 8 \, w \, \Box L.  
\ena 
This concludes our discussion of the vector-tensor multiplet in the new
context of 2-form central charge geometry. The invariant superfield 
and component field actions will be presented below after having included
coupling to Chern-Simons forms.

\sect{2-form geometry and Chern-Simons forms}

\indent

In order to prepare the ground for the coupling of Chern-Simons forms we 
return now to the more general setting of section {\bf 2}, \ie without
reality conditions as to the $z$, $\bz$ dependence. We consider a
Yang-Mills theory gauge potential $  \ \ca \ = \ E^\ca \ca_\ca \ $ 
together with its fieldstrength 
\be  
\cf \ = \ d \ca + \ca \ca \ = \ \frac{1}{2} E^\cb E^\cc \, \cf_{\cc \cb} 
    \ = \  E^\cb E^\cc \left( D_\cc \ca_\cb 
       - \ca_\cc \ca_\cb + \frac{1}{2} T_{\cc\cb} {}^\ca \ca_\ca \right),
\ee
satisfying Bianchi identities $ \ \cd \cf \ = \  0 \ $, in some more detail
\be E^\ca E^\cb E^\cc \left( \cd_\cc \cf_{\cb \ca} 
        + T_{\cc\cb} {}^\cf  \cf_{\cf \ca} \right) \ = \ 0. \ee
As is well-known, the covariant constraints 
\be  \cf {}_\bt^\tb {}_\al^\ta \ = \ 0, \hspace{1cm} 
     \cf {}_\tb^\db {}_\ta^\da \ = \ 0, \hspace{1cm} 
     \cf {}_\bt^\tb {}_\ta^\da \ = \ 0, \ee
allow to express the components of $\cf$ entirely in terms of the basic
spinorial gaugino superfields $\cw {}_\al^\ta$ and $\cwb {}_\ta^\da$.
They are identified in
\be \cf {}_\bt^\tb {\,}_a \ = \ +i  (\si_a \eps)_\bt {}^\db \, \cwb {}_\db^\tb, 
\hspace{1cm} 
\cf {}_\tb^\db {\,}_a \ = \ -i (\sib_a \eps)^\db {}_\bt \, \cw {}_\tb^\bt,\ee
and appear as well in 
\be
c^z \cf_z {}_\tb^\db \ = \ -4 \, \cwb {}_\tb^\db, \hspace{1cm}
\bc^\bz \cf_\bz {}_\bt^\tb \ = \ +4 \, \cw {}_\bt^\tb.  
\ee
The other nonvanishing components of the fieldstrength, as well as
the derivative constraints on $\cw {}_\al^\ta$ and $\cwb {}_\ta^\da$,
are summarized in  
\bea
\cd {}_\bt^\tb \, \cw {}_\al^\ta &=& 
    + \eps^{\tb \ta} (\si^{b a} \eps)_{\bt \al} \, \cf_{ba}
    + \eps_{\bt \al} \sum^{\tb \ta} \cd {}_\dv^{\tb} \cwb {}^{\dv \ta}  
    - \frac{1}{8} \eps_{\bt \al} \eps^{\tb \ta} c^z \bc^\bz \cf_{\bz z}, \\[2mm]
\cd {}_\tb^\db \, \cwb {}^\da_\ta &=&
    - \eps_{\tb \ta} (\sib^{b a} \eps)^{\db \da} \, \cf_{ba}
    + \eps^{\db \da} \sum_{\tb \ta}\cd {}^\vp _{\tb} \cw {}_{\vp \ta}
    + \frac{1}{8} \eps^{\db \da} \eps_{\tb \ta} \bc^\bz c^z \cf_{z \bz}.  
\ena
Moreover, one has
\be 
\cd {}_\bt^\tb \, \cwb {}^\da_\ta \ = \ 
   - \frac{i}{2} \dt^\tb_\ta \, (\si^a \eps)_\bt {}^\da \ c^z \cf_{\! z \, a},
\hspace{1cm}
\cd {}_\tb^\db \, \cw {}_\al^\ta \ = \
   + \frac{i}{2} \dt^\ta_\tb \, 
             (\sib^a \eps)^\db {}_\al \ \bc^\bz \cf_{\! \bz \, a}.  
\ee 
Having established the general scenario for the Yang-Mills gauge structure
we shall now define the corresponding Chern-Simons forms as
\be \cq \ = \ \tr (\ca \cf - \f{1}{3} \ca \ca \ca), \ee
which are coupled to the 2-form geometry 
as described in section {\bf 2} according to
\be \ch \ = \ dB + k \cq. \ee
The modified Bianchi-identities $d \ch = k \, \tr(\cf \cf)$ for 
this case result
in a number of modifications in the components of $H_{\cc \cb \ca}$ and the
constraints on $L$ compared to the results in section {\bf 2}. One finds in
particular
\bea
c^z \ch_z {\,}_{ba} &=& 
      (\eps \si_{ba})^{\bt \al} D_{\bt \ta} D {}^\ta_\al L
      - 2k \,  (\eps \sib_{ba})_{\db \da} \, 
            \tr \left( \cwb^{\db \ta} \cwb {}_\ta^\da \right), \\
\bc^\bz \ch_\bz {\,}_{ba} &=& 
      (\eps \sib_{ba})_{\db \da} \, D^{\db \ta} D {}_\ta^\da L
       -2k \, (\eps \si_{ba})^{\bt \al} \, 
            \tr \left( \cw_{\bt \ta} \cw {}^\ta_\al \right),
\ena
as well as
\be [ D {}_\bt^\tb,  D {}_\ta^\da ] L \ = \ 
              - 2 \, \dt^\tb_\ta \, \hat{h}_\bt {}^\da 
              - 4k \, \tr \left( \cw {}^\tb_\bt \cwb {}_\ta^\da \right), \ee
with
\be \hat{h}_\bt {}^\da \ = \ h_\bt {}^\da 
              - 2k \, \tr \left( \cw {}^\ta_\bt \cwb {}_\ta^\da \right). \ee
Furthermore
\bea 
\sum_{\tb \ta} D_{\da \tb} D {}_\ta^\da L 
        &=& 4k \, \tr \left( \cw {}^\al_\tb  \cw {}_{\al \ta} \right), \\
\sum^{\tb \ta} D^{\al \tb} D {}^\ta_\al L 
        &=& 4k \, \tr \left( \cwb {}^\tb_\da \cwb {}^{\da \ta} \right).
\ena
Finally, 
\be c^z \bc^\bz H_{\bz \, z \ a} \ = \ 8i \, \hat{h}_a, \ee
while the components $H_{\al \, z \bz}^{\ta}$ and $H_{\ta \ z \bz}^{\da}$
are given as before.
 
\sect{Vector-tensor multiplet and coupling to Chern-Simons forms}

\indent

The Yang-Mills multiplet of ref.\cite{GSW78} is recovered from
the more general geometric structure described in the previous section
in requiring all superfields (in the Yang-Mills sector) 
to be independent of $z$ and $\bz$. This means in particular that
the Yang-Mills gauge potentials having vanishing Lie derivatives
in the directions of $c^z$ and $\bc^\bz$,
\be L_{c^z} \ca \ = \ L_{\bc^\bz} \ca \ = \ 0, \ee
and similarly for the gauge parameter superfields.
As a consequence, the components $\ca_z$ and $\ca_\bz$ become gauge
covariant superfields.
Written in covariant form the same conditions become
\be \imath_{c^z} \cf \ = \ - D \bar{X}, \hspace{1cm} 
      \imath_{\bc^\bz} \cf \ = \ - D X, \ee
where we have defined $c^z A_z = \bar{X}$ and $\bc^\bz A_\bz = X$.
More explicitly,  one obtains
the familiar relations\footnote{The superspace geometry 
of \cite{GSW78} is then easily obtained
from an appropriate adjustement of conventional constraints as
explained in \cite{S78,G96}.}
\be 
\cd {}_\al^\ta \bar{X} \ = \ 0, \hspace{1cm}
\cd {}_\al^\ta X \ = \ - 4 \, \cw {}_\al^\ta,  
\ee
\be
\cd {}_\ta^\da \bar{X} \ = \ + 4 \, \cwb {}_\ta^\da, \hspace{1cm}
\cd {}_\ta^\da X \ = \ 0.  
\ee

Having specified the Yang-Mills gauge structure, whose Chern-Simons
form is to be coupled to the 2-form of the vector-tensor multiplet,
we have now to reconsider the identification of the abelian vector
gauge potential in the presence of Chern-Simons forms.

In order to get some intuition we shall consider first infinitesimal
Yang-Mills gauge transformations which change the gauge potential
$\ca$ and its Chern-Simons form $\cq$ according to
\be 
\dt \ca \ = \ - d \al - [\al, \ca],
\hspace{1cm} \mbox{and} \hspace{1cm}
\dt \cq \ = \ d \, \tr (\ca \, d \al). 
\ee
The variation of the Chern-Simons form in
the fieldstrength $\ch = dB + k \cq$ can be compensated
by assigning the transformation law
\be \dt B \ = \ d \bt - k \, \tr (\ca \, d \al) \ee
to the 2-form gauge potential. Given this modified transformation law
we define the 1-forms
\be 
\cv \ = \ \imath_{c^z} B + k \, \tr (\ca \, \imath_{c^z} \ca), 
\hspace{1cm}
\bar{\cv} \ = \ \imath_{\bc^\bz} B + k \, \tr (\ca \, \imath_{\bc^\bz} \ca),
\ee
subject to gauge transformations
\be 
\dt \cv \ = \ L_{c^z} \bt - d \, \imath_{c^z}\bt,  
\hspace{1cm}
\dt \bar{\cv} \ = \ L_{\bc^\bz} \bt - d \, \imath_{\bc^\bz} \bt.
\ee
Requiring $L_{c^z} \bt = L_{\bc^\bz} \bt$ as well as
$L_{c^z} B = L_{\bc^\bz} B$ we define then the abelian 1-form
gauge potential $A = \bar{\cv} - \cv$, whose fieldstrength
\be 
F \ = \ d A \ = \ \imath_{c^z} \ch - \imath_{\bc^\bz} \ch
      - 2k \, \tr ((\bar{X}-X) \cf),
\ee
is invariant under the gauge transformations 
$\dt A = d (\imath_{c^z}\bt - \imath_{\bc^\bz} \bt)$. 
As a consequence of their definition, the fieldstrength components
$F_{\cb \ca}$ are completely expressed in terms of the superfields $L$
and $X$, $\bar{X}$. In particular, the non-vanishing components
at canonical dimensions $0$ and $1/2$ are given as
\be
F{}_\bt^\tb {}_\al^\ta \ = \ +8 \eps_{\bt \al} \eps^{\tb \ta} L, \hspace{1cm} 
     F{}_\tb^\db {}_\ta^\da \ = \ -8 \eps^{\db \da} \eps_{\tb \ta} L,
\ee
\be 
F{}_\bt^\tb {\,}_a \ = \ +i  (\si_a \eps)_\bt {}^\db \, \bar{\Gamma} {}_\db^\tb, 
\hspace{1cm} 
F{}_\tb^\db {\,}_a \ = \ -i (\sib_a \eps)^\db {}_\bt \, \Gamma {}_\tb^\bt,
\\[2mm] \ee
\be 
c^z F_z{\, }_\al^\ta 
   \ = \ - \bc^\bz F_\bz{\, }_\al^\ta + 4 \Gamma {}_\al^\ta 
   \ = \ -8 D {}_\al^\ta L, \hspace{1cm}
\bc^\bz F_\bz {\, }_\ta^\da 
   \ = \ -c^z F_z {\, }_\ta^\da -4 \bar{\Gamma} {}_\ta^\da
   \ = \ + 8 D {}_\ta^\da L.
\ee
Here, we have defined the superfield 
\be \Gamma = L + \frac{k}{16} \tr (\bar{X}-X)^2, \ee
together with its spinorial derivatives 
$\Gamma {}_\tb^\bt = -4 D {}_\tb^\bt \Gamma$ 
and $\bar{\Gamma} {}_\db^\tb = -4 D {}_\db^\tb \Gamma$.
Moreover, due this definition and the special properties of the superfields
$L$ and $X$, $\bar{X}$ induced from the 2-form and Yang-Mills geometries, 
respectively, one obtains
\be 
\sum^{\tb \ta} D^{\al \tb} D {}^\ta_\al \Gamma \ = \ 
\sum^{\tb \ta}  D {}^\tb_\da D^{\da \ta} \Gamma, 
\ee
as well as 
\be 
\sum^{\tb \ta} D {}^\tb_\db D {}^\ta_\al \Gamma \ = \ 0, \hspace{1cm}
\sum^{\tb \ta}  D {}^\ta_\al D {}^\tb_\db \Gamma \ = \ 0.
\ee 
Obviously, the remaining components of $F$, at canonical dimension 1,
are related to the superfield expansion of the basic superfields
$L$ and $X$, $\bar{X}$ as well, in particular
\be 
F_{ba} \ = \ - \frac{1}{4} (\si_{ba})_\al {}^\bt D_\bt^\tb \Gamma^\al_\tb
 + \frac{1}{4} (\sib_{ba})^\da {}_\db D_\tb^\db \bar{\Gamma}^\tb_\da.
\ee

With the complete geometric description of Chern-Simons couplings at hand
it is now straightforward to write down an
invariant action and to identify component fields together
with their supersymmetry and central charge transformations.

As to the construction of a supersymmetric action we define the superfield
\be \Sigma \ = \ L - \frac{k}{16} \tr (X+\bar{X})^2. \ee
It is easy to convince oneself that $\Sigma$, similarly
to $\Gamma$, has the properties
\be 
\sum^{\tb \ta} D^{\al \tb} D {}^\ta_\al \Sigma  
+\sum^{\tb \ta}  D {}^\tb_\da D^{\da \ta} \Sigma \ = \ 0, 
\ee
as well as 
\be 
\sum^{\tb \ta} D {}^\tb_\db D {}^\ta_\al \Sigma \ = \ 0, \hspace{1cm}
\sum^{\tb \ta}  D {}^\ta_\al D {}^\tb_\db \Sigma \ = \ 0.
\ee
As a consequence, the combination
\be 
M^{\stackrel{\tb \ta}{\smile}} \ = \  
\sum^{\tb \ta} \left( D^{\al \tb} \Sigma \ D{}^\ta_\al \Sigma
+ \Sigma D^{\al \tb} D {}^\ta_\al \Sigma
-D {}^\tb_\da \Sigma \ D^{\da \ta} \Sigma \right), 
\ee
satisfies
\be \oint_{\tc \tb \ta} 
D_{\underline{\gm} \tc} M_{\stackrel{\tb \ta}{\smile}} \ = \ 0, \ee
for $\underline{\gm} = \gm, \dg$, and can, therefore, be employed
for the construction of a supersymmetric component field action
\cite{HOW97},\cite{H97},
obtained as the lowest component of the superfield
\be 
\left( D^{\al \tb} D {}^\ta_\al - D {}^\tb_\da D^{\da \ta} \right)
M_{\stackrel{\tb \ta}{\smile}}.
\ee
A straightforward calculation leads then to the component field action 
(canonical normalization of kinetic terms is easily established in performing
suitable field redefinitions)
\bea
\lefteqn{\cl_{VT} \ = \ 
    - \frac{1}{2} \, \partial^m L \, \partial_m L
    + \frac{1}{2} \, \hat{h}^m \hat{h}_m   
    - \frac{1}{64} \, \Sigma^{mn} \Sigma_{mn} 
    + \frac{i}{2} \, (\si^m \eps)_\al {}^\da \, \Lambda^\al_\ta 
        \stackrel{\leftrightarrow}{\partial}_m \bar{\Lambda}_\da^\ta
    - \frac{1}{8} \, D^2}
\nn \\[3mm] && 
    + g_{\raisebox{-.4ex}{\tiny{(i)(j)}}} \left\{ 
      - \, \cd_m X^{\raisebox{.5ex}{\tiny{(i)}}} \, 
      \cd^m \bar{X}^{\raisebox{.5ex}{\tiny{(j)}}} 
    + 2 \, \cf_{mn}^{\raisebox{.5ex}{\tiny{(i)}}} \, 
      \cf^{{\raisebox{.5ex}{\tiny{(j)}}} \, mn} 
    - 4i \, (\si^m \eps)_\al {}^\da 
          \, \cw{\!}^{\raisebox{.5ex}{\tiny{(i)}}}{}^\al_\ta 
          \, \stackrel{\leftrightarrow}{\cd}_m 
          \bar{\cw}^{\raisebox{.5ex}{\tiny{(j)}}}{}_\da^\ta
\right. \nn \\[2mm] && \left. 
\hspace{1.3cm} 
  - \frac{1}{2} \, 
    {\bf D}^{\raisebox{1.2ex}{\tiny{(i)}} \, \stackrel{\tb \ta}{\smile}} \,  
    {\bf D}^{\raisebox{1.2ex}{\tiny{(j)}}}_{\stackrel{\tb \ta}{\smile}} 
      - \frac{1}{16} \, [X,\bar{X}]^{\raisebox{.5ex}{\tiny{(i)}}}
              \, [X,\bar{X}]^{\raisebox{.5ex}{\tiny{(j)}}}    
      + 2 \, \cw{\!}^{\raisebox{.5ex}{\tiny{(i)}}}{}^\al_\ta
         \, [\bar{X}, \cw^\ta_\al]^{\raisebox{.5ex}{\tiny{(j)}}} 
      - 2 \, \cwb{\!}^{\raisebox{.5ex}{\tiny{(i)}}}{}_\da^\ta
         \, [X, \cwb_\ta^\da]^{\raisebox{.5ex}{\tiny{(j)}}}  \right\}
\nn \\[2mm] &&
    + \frac{k^2}{2} \, \tr \left( (X + \bar{X}) \cw^\al_\ta \right)
       \si^{mn}{}_\al {}^\bt \, \tr \left( \cw_\bt^\ta \cf_{mn} \right)  
    + \frac{k^2}{2} \, \tr \left( (X + \bar{X}) \cwb_\da^\ta \right)
       \sib^{mn}{}^\da {}_\db \, \tr \left( \cwb^\db_\ta \cf_{mn} \right)  
\nn \\[2mm] &&
     + \frac{k^2}{8} \, \cw{\!}^{\raisebox{.5ex}{\tiny{(i)}}}{}^{\al \ta}
       \si^m_{\al \da} \cwb{\!}^{\raisebox{.5ex}{\tiny{(j)}}}{}^\da_\ta
       \, {\bf t}_m^{\stackrel{\mbox{\tiny{(i)}} \mbox{\tiny{(j)}}}{\smile}}
+ \frac{1}{2} \, g_{\raisebox{-.4ex}{\tiny{(i)(j)}}} \,  
    Q^{\raisebox{1.2ex}{\tiny{(i)}} \, \stackrel{\tb \ta}{\smile}} \,  
    Q^{\raisebox{1.2ex}{\tiny{(j)}}}_{\stackrel{\tb \ta}{\smile}}
+ \frac{k^2}{4}  
       q^{\stackrel{\tb \ta}{\smile}} q_{\stackrel{\tb \ta}{\smile}}. 
\nn \\[2mm] &&
         + \frac{ik}{8} \, \hat{h}^m 
              \, \tr \left( (X + \bar{X}) \cd_m (X - \bar{X}) \right)
         + \frac{ik}{32} \, \vep^{klmn} \Sigma_{kl} 
             \tr \left( (X + \bar{X}) \cf_{mn} \right)  
\nn \\[2mm] &&
+k \Lambda^\al_\ta \, \tr \! \left(
\si^{mn}{}_\al {}^\bt \, \cw_\bt^\ta \cf_{mn}
- \frac{1}{8} [X,\bar{X}] \, \cw^\ta_\al 
- \frac{i}{2} \cwb^{\da \ta} \cd_{\al \da} X
\right) 
\nn \\[2mm] &&
-k \bar{\Lambda}_\da^\ta \, \tr \! \left(
\sib^{mn}{}^\da {}_\db \, \cwb^\db_\ta \cf_{mn}
- \frac{1}{8} [X,\bar{X}] \, \cwb^\da_\ta
- \frac{i}{2} \cwb_{\al \ta} \cd^{\da \al} \bar{X}
\right) 
\ena
We have used the convention 
$\tr(AB) = A^{\mbox{\tiny {(i)}}} B^{\mbox{\tiny{(i)}}}$
and the abbreviations
\be 
{\bf t}_m^{\stackrel{\mbox{\tiny{(i)}} \mbox{\tiny{(j)}}}{\smile}} \ = \
i \sum^{\mbox{\tiny{(i)}} \mbox{\tiny{(j)}}} 
        (X + \bar{X})^{\raisebox{.5ex}{\tiny{(i)}}} \, \cd_m
        (X - \bar{X})^{\raisebox{.5ex}{\tiny{(j)}}},
\ee
as well as
\be q_{\stackrel{\tb \ta}{\smile}} \ = \ \tr \! \left(
\cwb_{\da \tb} \cwb^\da_\ta - \cw^\al_\tb \cw_{\al \ta} \right), \ee
and
\be Q^{\raisebox{.5ex}{\tiny{(i)}}}_{\stackrel{\tb \ta}{\smile}} \ = \
\bar{g}^{\raisebox{.4ex}{\tiny{(i)(j)}}} \, \left(
\frac{k^2}{8} (X + \bar{X})^{\raisebox{.5ex}{\tiny{(j)}}}
       q_{\stackrel{\tb \ta}{\smile}}
- \frac{k}{4} \sum_{\tb \ta} \left( 
     \Sigma^\al_\tb \, \cw{\!}^{\raisebox{.5ex}{\tiny{(j)}}}_{\al \ta}
   + \bar{\Sigma}_{\da \tb} \cwb^{\raisebox{.5ex}{\tiny{(j)}}}{}^\da_\ta 
\right) \right), \ee
with $\bar{g}^{\raisebox{.4ex}{\tiny{(i)(j)}}}$ the inverse of the
gauge coupling function 
\be 
g_{\raisebox{-.4ex}{\tiny{(i)(j)}}} \ = \
  \frac{k}{4} \left( \Sigma \, \delta_{\raisebox{-.4ex}{\tiny{(i)(j)}}}
   - \frac{k}{8} \, (X + \bar{X})_{\raisebox{-.4ex}{\tiny{(i)}}}
                    (X + \bar{X})_{\raisebox{-.4ex}{\tiny{(j)}}} \right) 
\ee
and
\be 
\Sigma_\al^\ta \ = \ D_\al^\ta \Sigma, \hspace{1cm}
\bar{\Sigma}^\da_\ta \ = \ D^\da_\ta \Sigma.
\ee
Of course, for
$k=0$, one obtains simply the usual free theory for the pure
vector-tensor multiplet, which describes a non-interacting theory.
Interestingly enough, the coupling to Chern-Simons forms already
in itself induces a kinetic term for the Yang-Mills sector, with
field-dependent gauge coupling function $g_{\raisebox{-.4ex}{\tiny{(i)(j)}}}$. 
Other $k$-dependent modifications are encoded in the definitions
\be 
\hat{h}^k \ = \ \vep^{klmn} \left( \f12 \partial_l B_{mn} 
+  k \; \tr(\ca_l \partial_m \ca_n - \f23 \ca_l \ca_m \ca_n ) \right)
+ k \, \sib^{k \, \da \al} \, \tr \! \left( \cw_\al^\ta \cwb_{\da \ta} \right)
\ee
and
\be
\Sigma_{mn} \ = \ F_{mn} -2k \; \tr \! \left((X - \bar{X})\cf_{mn}
  -2 \, \cw^\al_\ta (\si_{mn})_\al {}^\bt \cw^\ta_\bt 
  +2 \, \bar{\cwb}_\da^\ta (\sib_{mn})^\da{}_\db \bar{\cwb}^\db_\ta \right),
\ee
with the field tensor  $F_{mn} \ = \ \partial_m A_n - \partial_n A_m $
identified in the geometric description given above.
As to the definitions of the component fields 
$X$, $\bar{X}$ and $\cw^\al_\ta$, $\cwb_\da^\ta$ 
in the Yang-Mills sector
we have deliberately used the same symbols for component fields,
identified as lowest component of the corresponding superfields.
Moreover, we have
\be
\cf_{mn} \ = \ \partial_m \ca_n - \partial_n \ca_m + [\ca_m,\ca_n],  
\ee
and
\be
{\bf D}^{\raisebox{.5ex}{\tiny{(i)}}}_{\stackrel{\tb \ta}{\smile}} \ = \   
\frac{1}{2} \, \sum_{\tb \ta} 
\cd^\al_\tb \cd_{\al \ta} X^{\raisebox{.5ex}{\tiny{(i)}}}
+ Q^{\raisebox{.5ex}{\tiny{(i)}}}_{\stackrel{\tb \ta}{\smile}}.
\ee
Supersymmetry and central charge transformations of the component fields
which leave the above lagrangian invariant are easily deduced from
the underlying geometric framework in superspace and will not be
rephrased here explicitly.
\sect{Conclusion and outlook}

\indent

We have shown that the vector-tensor multiplet can be obtained from
the two-form geometry in central charge superspace with suitably
chosen constraints. This formulation allows to derive central charge
transformations of the component fields a priori on purely
geometric grounds. Moreover, from this point of view, the analogies
of the vector-tensor multiplet with the linear multiplet of $N=1$
supersymmetry become quite evident. This analogy can be pursued
in order to construct an interacting theory due to Chern-Simons coupling.
Finally, we have strong reasons to believe that this formulation
should be quite helpful in the coupling of the vector-tensor
multiplet to supergravity and the study of nontrivial
central charges in the framework of local supersymmetry.

\end{document}